\documentclass[twocolumn,superscriptaddress,aps,preprintnumbers,amsmath,amssymb,prl,nofootinbib,10pt]{revtex4-2}

\usepackage{graphicx} 
\usepackage{dcolumn}
\usepackage{bm}
\usepackage[T1]{fontenc}
\usepackage{booktabs}
\usepackage{bbold}
\usepackage{mathtools}
\usepackage{multirow}
\usepackage{physics}
\usepackage{slashed}
\usepackage{colortbl}
\usepackage{hyperref}
\usepackage[dvipsnames]{xcolor}
\usepackage{orcidlink}
\usepackage{tikz,xcolor,hyperref}
\usepackage{adjustbox}

\graphicspath{{Figures/}}
\usepackage[compat=1.1.0]{tikz-feynman}

\newcommand\myshade{85}
\colorlet{mylinkcolor}{Blue}
\colorlet{mycitecolor}{Blue}
\colorlet{myurlcolor}{NavyBlue}
\hypersetup{
     linkcolor  = mylinkcolor,
     citecolor  = mycitecolor!\myshade!black,
     urlcolor   = myurlcolor!\myshade!black,
     colorlinks = true,
}

\definecolor{Gray}{gray}{0.85}
\definecolor{LightCyan}{rgb}{0.88,1,1}
\newcolumntype{a}{>{\columncolor{Gray}}c}

\DeclareMathOperator{\GeV}{\text{GeV}}
\DeclareMathOperator{\TeV}{\text{TeV}}

\DeclareMathOperator{\Lag}{\mathcal{L}}

\begin{document}

\title{Displaced vertex signals of low temperature baryogenesis}

\author{Pedro Bittar \orcidlink{0000-0002-3684-5692}}
\email{bittar@if.usp.br}

\author{Gustavo Burdman \orcidlink{0000-0003-4461-2140}}
\email{gaburdman@usp.br}

\affiliation{Department of Mathematical Physics,
    Institute of Physics, 
    University of São Paulo, \\
    R. do Matão 1371, São Paulo, 
    SP 05508-090, Brazil
}

\date{\today}

\begin{abstract}
\noindent We explore the connection of baryogenesis at temperatures below the electroweak scale and signals for long-lived particles at the LHC. The model features new SM singlets, with a long-lived fermion decaying to quarks to generate the baryon asymmetry. The model avoids strong flavor physics bounds while predicting a rich diquark phenomenology, monojet signals, and displaced vertices. We show how the displaced vertex signals can be probed at the HL-LHC. The large transverse production makes a strong physics case for constructing far detector experiments such as MATHUSLA, ANUBIS, and CODEX-b,  complementary to the central and forward long-lived particle program.
\end{abstract}

\maketitle
\noindent\textbf{Introduction\,\,---\,\,}
The observed matter anti--matter asymmetry remains one of the biggest motivations for physics beyond the standard model (BSM). Some of the traditional BSM baryogenesis mechanisms such as electroweak baryogenesis \cite{Cohen:1993nk,Wagner:2023vqw}, leptogenesis \cite{Davidson:2008bu}, and Affleck-Dine models \cite{Allahverdi:2012ju} generate the baryon asymmetry at high temperatures, above the electroweak (EW) scale, and sometimes much above, e.g. $\sim 10^{15} \GeV$. While these are well-motivated models, they can be elusive for testing given the high new physics scales involved. Alternatively, low-temperature baryogenesis models propose to generate the baryon asymmetry after the EW transition \cite{Dimopoulos:1987rk,Trodden:1999ie,Babu:2006xc,Babu:2006wz,Kohri:2009ka,Allahverdi:2010im,Allahverdi:2022zqr}. Baryogenesis occurs through a new particle decaying after the sphaleron decoupling and before big bang nucleosynthesis (BBN) times. These decays violate baryon number and generate a CP asymmetry, meeting all three Sahkarov conditions \cite{Sakharov:1967dj}.

If a particle $N$ decays to generate the asymmetry, its decay length has to satisfy the out-of-equilibrium condition, $\tau_N > H^{-1}$ at temperatures $T\simeq m_N$. For masses around the EW scale, this relation implies 
\begin{equation}
    c\tau_N > 20 ~\text{mm} \left(m_N/100 \GeV\right)^{-2},
\end{equation}
resulting in a decay length of macroscopic size. Furthermore, if the lifetime of $N$ is longer than a picosecond, its decays occur after the Sphaleron transition. If it is shorter than a second the decay occur before BBN. This connection between long-lived particles (LLP) and low-temperature baryogenesis represents an excellent opportunity for displaced vertex searches precisely because they select the decay length window of $0.1 \text{mm} \lesssim c \tau_N \lesssim  200 \text{m}$. Such signals can be most effectively searched at the Large Hadron Collider (LHC) and dedicated far detector experiments such as  MATHUSLA \cite{MATHUSLA:2020uve,Curtin:2018mvb}, CODEX-b \cite{Aielli:2022awh,Gligorov:2017nwh}, ANUBIS \cite{Bauer:2019vqk}, AL3X \cite{Gligorov:2018vkc} and the Forward Physics Facilities \cite{Feng:2022inv} with FASER/FASER2. 
\begin{figure}
    \centering
    \includegraphics[scale=0.9]{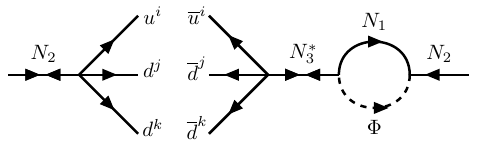}
    \caption{Diagrams for CP violation in the interference between tree-level and loop decays of $N_2$ with and without mixing with the heavier state $N_3$. The lightest singlet fermion $N_1$ and scalar $\Phi$ are on-shell in the loop to generate a CP asymmetry.}
    \label{fig:CP_violation}
\end{figure}
The relationship between LLP and baryogenesis models is surprisingly under-explored in the literature, with only a few proposed models. As reviewed in \cite{Curtin:2018mvb}, some of these models are WIMP baryogenesis \cite{Cui:2012jh,Cui:2013bta,Cui:2014twa,Cui:2016rqt} and baryogenesis via exotic baryon oscillations \cite{Ipek:2016bpf,Aitken:2017wie,McKeen:2015cuz}. In these, charged BSM states enter on-shell in the decay loop amplitude so as to generate CP violation. Because extra SM-charged particles should be heavy, such models prefer masses above the EW scale. Instead, we focus on a lower mass range, $10\GeV \gtrsim m_N \gtrsim 200 \GeV$, resulting in generally longer lifetimes. To reach these masses, we propose a model with only SM singlets below the TeV scale, one of which can successfully decay to create a baryon asymmetry. We combine several features of different models proposed in the literature \cite{Davoudiasl:2010am,Davoudiasl:2015jja,Arnold:2012sd,Assad:2017iib,Cheung:2013hza,Allahverdi:2010rh,Allahverdi:2013mza,Allahverdi:2017edd,Elor:2022hpa}. 
The resulting simplified framework naturally avoids proton decay, neutron oscillations, neutron electric dipole moment (EDM), and strong flavor physics bounds while providing a diverse phenomenology for colliders and cosmology. In a companion paper \cite{Bittar:2024nrn}, we expand the model to include spontaneous symmetry breaking of baryon number. Here we focus on the model's aspects that are important for displaced vertices signals.

\vspace{1mm}\noindent\textbf{Model\,\,---\,\,}%
We consider three flavors of singlet Majorana fermions, $N_\alpha=N_{{1,2,3}}$, one neutral scalar, $\Phi$, along with new dynamics at a high UV scale. Baryon number violation fixes the dynamics at the UV scale, which generates an effective coupling between $N$ and a neutral combination of quarks. The simplest operator that couples $N$ to SM baryon number happens at dimension six. Since $N$ is neutral, the only quark flavor structure allowed for pairing is $udd'$. A minimal possibility that meets these requirements is
\begin{align}
    \Lag_{\rm eff}=\frac{\kappa_{\alpha}^{ijk}}{M_X^2}\big(\overline{N_\alpha^c} u_R^i \big)\big(\overline{{d_R^c}}^j d_R^k\big) +\xi_{\alpha\beta}\,\overline{N_\alpha^c}\Phi N_\beta + h.c.
    \label{eq:Leff}
\end{align}

\noindent Here, $M_X$ is the UV scale, $i,j,k$ are quark generation indices and $\alpha,\beta$ are the neutral fermion flavors. We assume a mass hierarchy of $m_{N_1}\lesssim m_\Phi < m_{N_2}<m_{N_3}$. Then, CP violation occurs by the interference between decays with and without mixing of one of the neutral fermions and another heavier state as shown in FIG.~\ref{fig:CP_violation}. A non-zero CP phase requires the mixing term to have on-shell intermediate states in the loop. Therefore, the intermediate states must be lighter than the decaying particle. These requirements show that CP violation occurs exclusively in the decay of $N_2$ as it mixes at loop-level with the heavier $N_3$ and has an on-shell loop contribution from the lighter $N_1$.

The baryon asymmetry parameter is the product of the yield of $N_2$, $Y_2=\tfrac{n_{N_2}}{s}$, the CP asymmetry  $\epsilon_{CP}$ and the branching ratio of $N_2$ to quarks, $\text{Br}(udd')$.
\begin{equation}
Y_{\Delta B}= Y_2 \epsilon_{CP} \,\text{Br}(udd').
\label{eq:BasymN2}
\end{equation}

\noindent Using the diagrams of FIG. \ref{fig:CP_violation}, the CP asymmetry is computed to be
\begin{align}
    {\epsilon}_{CP}\simeq \frac{3}{8\pi}\frac{m_{N_1}}{m_{N_3}}\frac{|\kappa_2 \xi_{12}^* \,\xi_{13} \kappa_3^*|\sin\delta }{|\kappa_2|^2}\sqrt{1-\tfrac{m_{N_1}^2+m_\delta^2}{m_{N_2}^2}} ~,
    \label{eq:epsN2}
\end{align}

\noindent where $\delta$ is the phase resulting from the coupling product $\kappa_2^{ijk} \xi_{12}^*\xi_{13}\kappa_3^{ijk*}$. We summed over the quark final states $c,d,s,b$ assuming the flavor hierarchies described in the appendix, which allows us to write $\kappa_\alpha^{cjk}\approx \kappa_\alpha$ in (\ref{eq:epsN2}).

To compute $N_2$'s yield, we consider the thermal history of the neutral sector. The relevant processes are the annihilation of $N_2$ to quarks, $N_2 u \leftrightarrow d d'$, and the decay of $N_2 \rightarrow udd'$. For $m_{N_2} \sim 100 \GeV$ with small non-renormalizable interactions, the freeze-out of the annihilation processes can occur when $N_2$ is still relativistic, with freeze-out temperature $m_{N_2}<T_{FO} < M_{X}$. We can estimate the relativistic freeze-out temperature by setting $n \langle\sigma_{\rm ann} v \rangle \simeq H(T_{FO})$ during radiation domination to get,
\begin{equation}
    T_{FO}\simeq 280 \GeV \left(\frac{M_X}{2\TeV}\right)^{4/3}\left(\frac{10^{-6}}{\kappa_2}\right)^{2/3}
\end{equation}

\noindent The relativistic freeze-out of $N_2$ is welcome since it maximizes the baryon asymmetry by not having a Boltzmann suppressed population of $N_2$. With a sufficiently large lifetime, $N_2$ can decay after freeze-out and also after the sphaleron decoupling. We can obtain the $N_2$ yield using the relativistic equilibrium expression
\begin{equation}
    Y_2=\frac{45\zeta(3)}{2\pi^4}\frac{g_{N_2}}{g_{*,s}(T_{\rm FO})}~,
    \label{eq:YN2}
\end{equation}
where $g_{N_2}$ is the number of $N_2$ degrees of freedom,  and  $g_*(T_{FO})$ is the total number of degrees of freedom in the bath at the freeze out temperature.

Lastly, to compute the branching ratio, we must consider the relative contributions of the tree-body and two-body decays, $N_2\rightarrow udd'$ and $N_2\rightarrow N_1 \Phi$. For $m_{N_1}\sim m_\Phi$ the partial decay widths are given by
\begin{align}
    \Gamma_{N_2\rightarrow udd'} \simeq  \frac{3|\kappa_2|^2 }{192 \pi ^3}\frac{m_{N_2}^5}{M_X^4}, 
    \hspace{0.4cm}
    \Gamma_{N_2\rightarrow N_1 \Phi} \simeq  \frac{m_{N_2}|\xi_{12}|^2}{\pi}.
\end{align}

\begin{figure}
    \centering
    \includegraphics[scale=0.38]{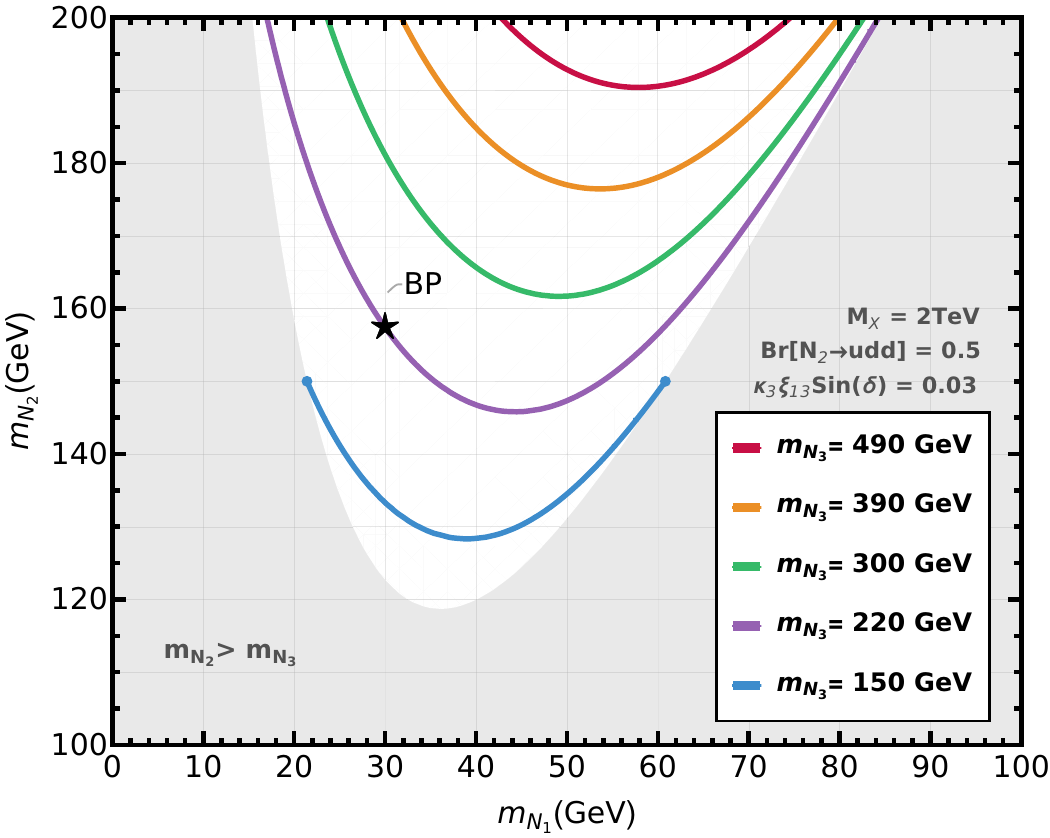}
    \caption{Masses of the particles in the model that reproduce the observed baryon asymmetry. We plot the curves respecting relation \eqref{eq:YB} in the $m_{N_1} \times m_{N_2}$ plane. Each curve represents a choice of mass $m_{N_3}$ for the scale $M_X=2\TeV$. We choose the values $\kappa_3 \xi_{13} \sin\delta=0.03$. We assume that $N_2$ decays to $u,d,s,c,b$ quarks, with a branching ratio of $0.5$. BP indicates the benchmark point chosen throughout the paper.}
    \label{fig:masses}
\end{figure}

\noindent The decay rate of $N_2\rightarrow udd'$ should be  comparable to $N_2\rightarrow N_1 \Phi$ in order to generate the baryon asymmetry efficiently. In turn, this requirement translates into a relationship between the couplings $\kappa_2$ and $\xi_{12}$,
\begin{equation}
    \label{eq:kxi12_relation}
    \frac{|\xi_{12}|}{|\kappa_2|} \approx \frac{1}{8\sqrt{3}\pi}\left(\frac{m_{N_2}}{M_X}\right)^2,
\end{equation}
where the equality corresponds to a $50\%$ branching ratio to the $udd'$ final state. Putting together \eqref{eq:epsN2} and \eqref{eq:YN2} and requiring \eqref{eq:kxi12_relation} to get a $\mathcal{O}(1)$ branching ratio to the baryon number violating channel, the baryon asymmetry for $M_X=2\TeV$ is given by
\begin{equation}
    \frac{Y_{\Delta B}}{Y_{\Delta B}^{\rm exp}}\simeq \left(\frac{m_{N_2}}{100 \GeV}\right)^2 \frac{\kappa_3 \xi_{13} \sin\delta}{3\times 10^{-2}}  \frac{m_{N_1}}{m_{N_3}}\sqrt{1-\tfrac{m_{N_1}^2+m_\delta^2}{m_{N_2}^2}}
    \label{eq:YB}
\end{equation}

\noindent where we used the central value $Y_{\Delta B}^{\rm exp}=8.7\times 10^{-11}$ measured by Planck \cite{Planck:2018vyg}. In FIG. \ref{fig:masses}, we impose the measured baryon asymmetry to fix the masses of the $N_{1,2,3}$.

It is useful to define a simple tree-level UV completion for \eqref{eq:Leff} by integrating in a $(\mathbf{3},\mathbf{1})_{2/3}$ SM charged scalar diquark, $X$. The allowed gauge invariant terms are
\begin{equation}
    \label{eq:L2/3}\mathcal{L}_{X}= \lambda_{\alpha i} X^\dagger \overline{N^c_\alpha} u_R^i + \lambda'_{jk} X \overline{d_R^c}^j d_R^k + h.c.
\end{equation}

\noindent Such diquark has some interesting low-energy properties. First, if $m_{N_\alpha}> 1\GeV$ and $N$ does not mix with neutrinos, the proton does not decay \cite{Arnold:2012sd}. Second, QCD gauge invariance implies color antisymmetry, which means the  flavor antisymmetry of the $\overline{d_R^c}^j d_R^k$ quark couplings in (\ref{eq:L2/3}). Therefore, there are only three independent $\lambda'$ couplings, $\lambda'_{jk}=\epsilon_{jkl}\lambda'_l$. Because of this, there is no tree-level neutron oscillation. The lightest baryon that oscillates is $\Lambda_0$, which imposes weak constraints on the diquark mass \cite{Bittar:2024nrn}. Additionally, there are no tree-level $K-\overline{K}$ and $B-\overline{B}$ mixing. At one loop, neutral Kaon mixing must involve the $b$ quark. In the case of B-meson mixing, the loop must contain an $s$ quark. Then, if one of the couplings is small, e.g. $\lambda'_{bs}< \lambda'_{ds}, \lambda'_{db}$, the bounds from meson oscillations can be negligible while still allowing for order one diquark couplings. Neutron-antineutron oscillations are suppressed by two-loop and CKM contributions rendering neglibible bounds for this model. Regarding neutron EDM bounds, the new CP-violating phases only contribute at three-loops or higher. A detailed discussion of low-energy bounds is given in the appendix.



\smallskip\noindent\textbf{LLPs at the LHC\,\,---\,\,}%
Having defined the UV model, we now discuss the LHC sensitivity for the predicted LLPs. We start by considering the direct production bounds on the diquark that govern the necessary UV dynamics of the model. CMS \cite{CMS:2022usq,CMS:2018mts} and ATLAS \cite{ATLAS:2017jnp} conducted searches for non-resonant pair production of dijet resonances. Their benchmark model is the R-parity-violating supersymmetric top squark that decays to $ds$ quarks and can be directly associated with the diquark $X$ above. The leading limits from \cite{CMS:2022usq} excludes the $Y=2/3$ scalar diquark at $95\%$ confidence level between $0.50 \TeV$ and $0.77 \TeV$, with a $3.6(2.5)$ local(global) excess occurring at $0.95 \TeV$. There are also resonant searches for $X$ to a pair of jets conducted by CMS \cite{CMS:2018mgb,CMS:2019gwf} and ATLAS \cite{ATLAS:2019fgd} through their diquark couplings. Ref. \cite{Pascual-Dias:2020hxo} imposed diquark bounds from resonant production for different values of the coupling pair $(\lambda'_{ds},\lambda'_{sb})$. For a diquark of mass $M_X=2 \TeV$, the values below $(\lambda'_{ds},\lambda'_{sb})=(0.3,0.12)$ are allowed by both direct searches and the flavor constraints of meson oscillations.


\begin{figure*}
    \flushleft
    \hspace{10mm}\includegraphics[scale=0.63]{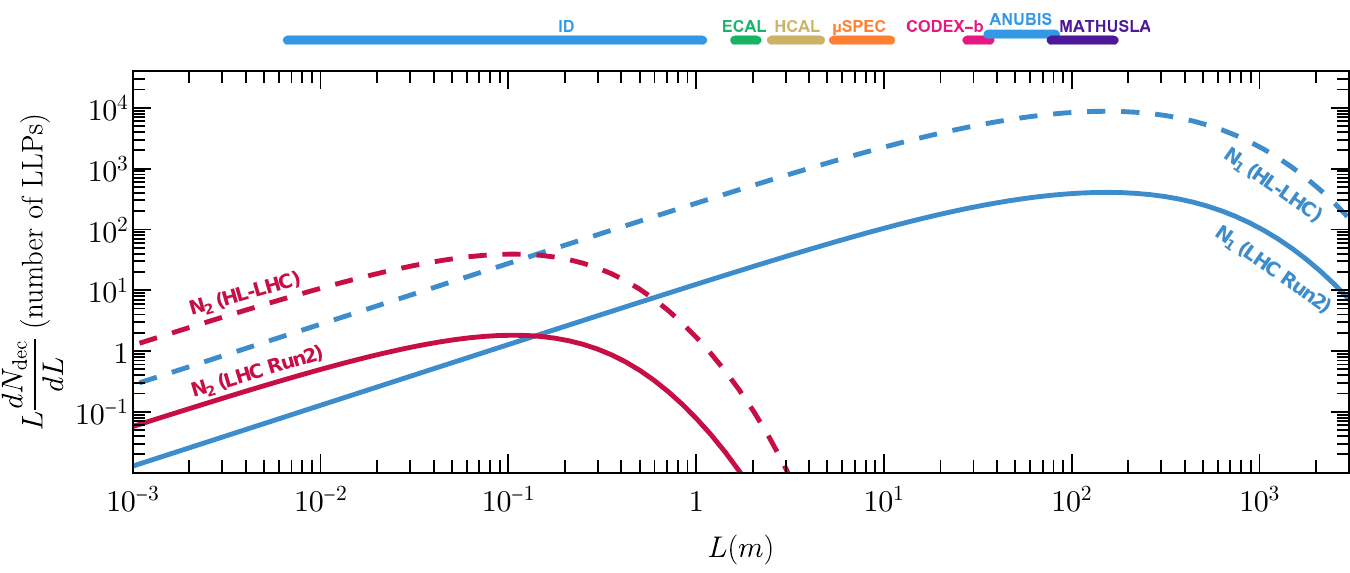}
    \caption{Differential distribution for the number of LLP decays as a function of the distance from the interaction point $L$. The red curve is the number of $N_2$ with proper decay length $c \tau_2 \simeq 1 \text{cm}$ and the blue is for $N_1$ with $c\tau_1 \simeq 10 \text{m}$. The solid lines are for LHC run 2 (R2) with $\mathcal{L}=139fb^{-1}$, and the dashed are for the high-luminosity phase (HL-LHC) with $\mathcal{L}=3000fb^{-1}$. The color lines above the plot indicate the ATLAS inner detector (ID), the electromagnetic (ECAL) and hadronic calorimeters (HCAL), the muon spectrometer ($\mu$SPEC) and the transverse far experiments CODEX-b, ANUBIS and MATHUSLA.}
    \label{fig:decaylength}
\end{figure*}

The production of $N_\alpha$ leads to monojet signals when there is a single $X$ or multi-jet plus missing $p_T$ for diquark pair production. However, pair production is subdominant for $M_X=2\TeV$ and couplings of $\lambda_{\alpha i}\lesssim 0.3$. Therefore we consider only the monojet channels in our analysis. Moreover, for the monojet channels, $N_\alpha$ can be singly produced in the $d s \rightarrow N u_i$ channel or doubly produced by the gluon initiated $g u \rightarrow X^{(*)} N \rightarrow N N' u_i$ and diquark $t$-channel $u \overline{u} (g) \rightarrow N N' (g)$ processes. In the case of the double production $t$-channel, the monojet comes from the initial state emission of a gluon, but these are limited by the high $p_T$ cut required by the analysis. For the other two channels, the resulting $p_T$ distribution of the jets is a Jacobian peak at $M_X/2$. Because of this feature, the cross-section can be sizable as it does not suffer from the high $p_T$ cuts. CMS \cite{CMS:2017tbk} and ATLAS \cite{ATLAS:2024xne} have searched for monojets in the context of fermion portal dark matter \cite{Bai:2013iqa} and light non-thermal dark matter models \cite{Dutta:2014kia}. In our model, these bounds are most relevant for $N_3$, which is produced and decays promptly to the LLPs. The monojet $N_3$ channel is the largest since $\kappa_3$ has to be maximized in \eqref{eq:BasymN2} to get a sizable $Y_{\Delta B}$. Because of this, the monojet $N_3$ channel is the one driving our LLP phenomenology, as we discuss next.
    
First we focus on the benchmark point shown in the plots with a star labeled \textit{BP}. We hold the diquark couplings fixed to $(\lambda'_{ds},\lambda'_{db},\lambda'_{sb})=(0.25,0.28,0.1)$. We choose $\lambda_{\alpha u}\ll\lambda_{\alpha c}=\lambda_{\alpha t}$ to respect the bounds described in the appendix and set $\lambda_{1c} = \lambda_{2c} = 5 \times 10^{-4}$ and $\lambda_{3c} = 0.1$, which gives decay lengths $c\tau_1 = 10 ~{\rm m}$ and $c\tau_2 = 1~\text{cm}$ for $N_1$ and $N_2$, while $N_3$ decays promptly. The production channel considered is $pp \rightarrow j N_3$, with $N_3$ decaying to either three $N_1$s or one $N_2$ and two $N_1$s. This happens through the $N_3\rightarrow N_{1,2} \Phi$ and $\Phi \rightarrow N_1 N_1$ cascade decay, where we assume the $\Phi$ Yukawa couplings to be $(\xi_{13},\xi_{23})=(1,0.3)$. For the signal simulation, we use \textsc{MadGraph\_aMC@NLO} version 2.9.19 with the parton distribution function \textsc{NNPDF2.3QED} and generate the UFO model using \textsc{FeynRules} \cite{Alloul:2013bka}. 

\begin{table*}[]
\begin{tabular}{|c|cccc|ccc|c|}
\hline
\multirow{2}{*}{\begin{tabular}[c]{@{}l@{}}\end{tabular}} &
  \multicolumn{4}{c|}{ATLAS/CMS} &
  \multicolumn{3}{c|}{Transverse Far Detectors} &
  Forward Far Detectors \\ \cline{2-9} 
 &
  \multicolumn{1}{c|}{ID$~(N_2)$} &
  \multicolumn{1}{c|}{ECAL} &
  \multicolumn{1}{c|}{HCAL} &
  \multicolumn{1}{c|}{$\mu$SPEC} &
  \multicolumn{1}{c|}{CODEX-b} &
  \multicolumn{1}{c|}{ANUBIS} &
  MATHUSLA &
  FASER/FASER2, ... \\ \hline 
$N_{\rm obs}$ (Run2) & \multicolumn{1}{c|}{$<3$} & \multicolumn{1}{c|}{$18\times \epsilon_{\rm recon}^{\rm LLP}$} & \multicolumn{1}{c|}{$116\times \epsilon_{\rm recon}^{\rm LLP}$} & \multicolumn{1}{c|}{$163\times \epsilon_{\rm recon}^{\rm LLP}$} & \multicolumn{1}{c|}{-} & \multicolumn{1}{c|}{-} & \multicolumn{1}{c|}{-} & \multicolumn{1}{c|}{<1} \\
$N_{\rm obs}$ (HL)   & \multicolumn{1}{c|}{$ 60$} & \multicolumn{1}{c|}{$397\times \epsilon_{\rm recon}^{\rm LLP}$} & \multicolumn{1}{c|}{$2509\times \epsilon_{\rm recon}^{\rm LLP}$} & \multicolumn{1}{c|}{$3537\times \epsilon_{\rm recon}^{\rm LLP}$} & \multicolumn{1}{c|}{$26$} & \multicolumn{1}{c|}{$72$} & \multicolumn{1}{c|}{$370$} & \multicolumn{1}{c|}{<1} \\ \hline
\end{tabular}
\caption{Number of observable displaced vertices from the decay of $N_{1,2}\rightarrow udd'$ inside the inner detector (ID), electromagnetic and hadronic calorimeters (ECAL and HCAL), and muon spectrometers($\mu$SPEC). We assume the large-tracking radius algorithm of \cite{ATLAS:2023oti} to estimate the efficiency of LLP reconstruction for decays inside the ID and optimistically choose $\epsilon_{\rm recon}^{\rm LLP}=1$ for far detectors. The number of events inside the ECAL/HCAL and $\mu$SPEC is left as a function of reconstruction efficiency.}
\label{tab:Nobs}
\end{table*}

We extract the angular and velocity distribution of $N_1$ and $N_2$ from the simulated data. The majority of events are transverse to the interaction plane, forming an approximately uniform angular distribution for $|\eta|<5$. From the velocity distribution, we extract the boosted lifetimes $\beta\gamma c\tau$, and construct the differential probability distribution for $N_{1,2}$ to decay at position $L$,
\begin{equation}
    \frac{d\mathcal{P}_{\rm dec}(L)}{dL} = \frac{e^{-\frac{L}{\beta\gamma c\tau_i}}}{\beta\gamma c\tau_i}.
    \label{eq:PDF_decay}
\end{equation}

\noindent Then, the differential number of events observed as a function of the distance $L$ is given by 
\begin{equation}
    \frac{dN_{\rm obs}}{dL}=\overbrace{(n\sigma\mathcal{L})\otimes\text{Br}(udd')\otimes\frac{d\mathcal{P}_{\rm dec}}{dL}}^{\equiv dN_{\rm dec}/dL}\otimes \xi_{geom}^{\rm LLP} \otimes \epsilon_{recon}^{LLP},
    \label{eq:dNobs/dL}
\end{equation}
\noindent where $n$ is the multiplicity of $N_{1,2}$ of the process, $\mathcal{L}$ is the luminosity, $\xi_{geom}^{LLP}$ is the geometric acceptance of the experiment and $\epsilon_{recon}^{LLP}$ is the efficiency of reconstruction of the displaced vertices. In FIG. \ref{fig:decaylength}, we obtain the differential number of LLP decays defined in the overbrace of \eqref{eq:dNobs/dL} as a function of the distance $L$. Notice that the value obtained is for the whole solid angle without specifying the geometric acceptance and detector efficiencies. The number of displaced vertices is sizable in several regions of the future LHC experiments, as well as for proposed far detectors with sensitivity to transverse events. 

\begin{figure}
    \flushleft \hspace{0.4cm}
    \includegraphics[scale=0.73]{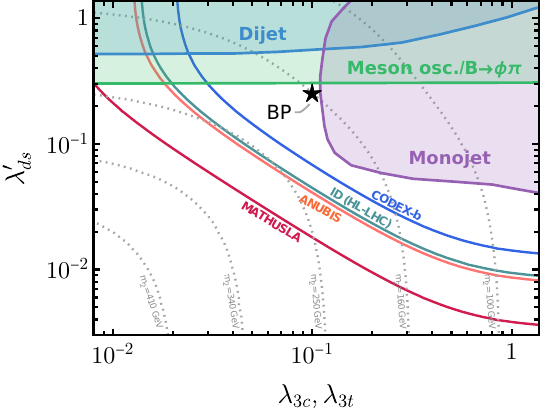}
    \caption{Exclusion bounds on $(\lambda_{3{c}}=\lambda_{3t},\lambda'_{ds})$. The purple region is excluded by CMS monojet searches \cite{CMS:2017tbk}, blue by ATLAS+CMS dijet searches, and the green by meson mixing constraints \cite{Giudice:2011ak,Pascual-Dias:2020hxo}. Dotted lines show the $m_{N_2}$ values required for the asymmetry \eqref{eq:YB}, assuming $m_{N_3} = m_{N_2} + 70\GeV$.
    }
    \label{fig:N3_bounds}
\end{figure}

To get an estimate of the number of LLPs in each experiment, we can integrate the length $L$ in \eqref{eq:PDF_decay} to obtain the decay probability inside a detector that starts at $L_{\rm in}$ and ends at $L_{\rm fin}$. In TABLE \ref{tab:Nobs}, we estimate the number of events for the benchmark point used throughout the paper for various LHC experiments during run 2 and the HL-LHC. For decays inside the inner detector of the ATLAS experiment, we assume the large-radius tracking reconstruction algorithm used in a similar search for RPV SUSY long-lived neutralinos \cite{ATLAS:2023oti}. Since the ATLAS RPV SUSY search targets heavy neutralinos, we only count $N_2$ decays inside the inner detector. A dedicated study of the three displaced vertices signal from $N_3$ decays into light $N_1$ plus a monojet could improve sensitivity but is left for future work. There are also displaced decays inside the calorimeters and muon spectrometers of the ATLAS/CMS experiments. However, we are not aware of a reconstruction strategy for the three-jet signals of the LLPs of our model beyond the ID. Because of that, we show the resulting number of events without specifying the reconstruction efficiencies of each part of the detector. For the far detectors, we consider the transverse and forward experiments CODEX-b, ANUBIS, MATHUSLA, and FASER/FASER2. To obtain the geometric acceptances, we integrate the number of events distribution over the geometric coverage of each experiment integrating from $L_{\rm in}$ to $L_{\rm fin}$ for each detector. We choose the optimistic scenario by assuming that the reconstruction efficiencies of the far detectors are $\epsilon_{\rm recon}^{\rm LLP}=1$. While the forward far detectors have negligible event rates due to the limited solid angle coverage, the transverse detectors present a significant sensitivity to  LLP decays. This enhanced sensitivity is primarily due to their optimal positioning relative to the interaction point, which allows for a larger geometric acceptance and increased event visibility. As shown in FIG. \ref{fig:decaylength}, the experiments considered are complementary in their range for probing $c\tau$. CODEX-b, ANUBIS and MATHUSLA are strategically positioned at successive distances from the interaction point, allowing for a continuous probing of the region ranging from $26$ to $170$ meters. Lastly, in Fig.~\ref{fig:N3_bounds}, we show the projected sensitivity of various HL-LHC experiments in the $(\lambda_{c3}, \lambda'_{ds})$ plane, assuming $\lambda_{c3} = \lambda_{t3}$ and $\lambda_{u3}\ll \lambda_{c3}$. In some regions, the displaced vertex searches extend the reach for this baryogenesis model by an order of magnitude compared to current flavor and monojet bounds. This highlights the potential of the HL-LHC to probe the low scale baryogenesis and motivates the construction of the proposed far detectors.


\smallskip\noindent\textbf{Conclusions\,\,---\,\,}%
In this paper, we explore the connection of baryogenesis through out-of-equilibrium decays at low temperatures and the detection of long-lived particles at the LHC. The proposed model can be probed through the detection of LLPs and associated phenomenology using current and future LHC experiments. The model requires the presence of three flavors of majorana fermions $N_{1,2,3}$, a neutral scalar $\Phi$, and some $\TeV$ scale dynamics which we assume to be related to a scalar diquark with $2/3$ hypercharge. The model avoids proton decay because $N$ does not mix with neutrinos, has no sizable neutron oscillations, and has small loop-suppressed $\Delta F = 2$ neutral meson oscillations. At the same time, there are plenty of interesting collider signals ranging from the detection of the diquarks to monojet signals with missing energy and displaced vertices, which are the main focus of the paper. The HL-LHC provides a promising scenario for probing the model as a significant number of LLPs can be produced. Transverse far detectors like CODEX-b, ANUBIS, and MATHUSLA have a significantly higher sensitivity to LLP decays compared to forward detectors due to their optimal positioning and larger geometric acceptance. The increased sensitivity of transverse far detectors highlights the importance of their construction and complementary ability to probe new physics at the HL-LHC. \\[0.5mm]

\begin{acknowledgments}
We thank Larissa Kiriliuk, Gabriel M. Salla, Lucas M. D. Ramos and Olcyr Sumensari for valuable discussions. Additionally, we acknowledge the financial support provided by FAPESP grant number 2019/04837-9 and CAPES grant number 88887.816450/2023-00.
\end{acknowledgments}

\bibliography{Bgen}
\bibliographystyle{JHEP}

\appendix
\begin{widetext}
\section{Supplementary Material: Low Energy Bounds}

 This appendix is dedicated to detailing the low energy bounds from the UV completed model with a scalar diquark $X=(\mathbf{3},\mathbf{1})_{2/3}$. In the end, we summarize the required couplings for the model to be compatible with current constraints. The Lagrangian of the model is 
\begin{equation}
    \label{eq:L2/3_app}\mathcal{L}= |\partial_\mu \Phi|^2 -V(\Phi) + \frac{1}{2} \overline{N}_{\alpha} (\slashed{\partial}- m_{N_\alpha}) N_{\alpha} - \xi_{\alpha\beta}\,\overline{N_\alpha^c}\Phi N_\beta + |D_\mu X|^ 2 - M_X^2 |X|^ 2 - \lambda_{\alpha i} X^\dagger \overline{N^c_\alpha} u_R^i - \lambda'_{jk} X \overline{d_R^c}^j d_R^k + h.c.
\end{equation}

\noindent As discussed in the main text, the QCD structure, together with the fermionic nature of the spinors, imply that the couplings $\lambda'_{jk}$ are antisymmetric in flavor. This means that there are three independent flavor combinations $\lambda'_{ds}, \lambda'_{db}$ and $\lambda'_{sb}$. When rotating to the mass basis, the coupling transforms as $
\lambda' \rightarrow \tilde{\lambda'} = U_{d_R}^T \lambda' U_{d_R}$. This transformation preserves the antisymmetry:
\begin{equation}
\tilde{\lambda'}^T = (U_{d_R}^T \lambda' U_{d_R})^T = U_{d_R}^T \lambda'^T U_{d_R} = -U_{d_R}^T \lambda' U_{d_R} = -\tilde{\lambda'}.
\end{equation}

\noindent Therefore, the coupling remains antisymmetric in the mass basis. This property holds as long as the coupling involves only down-type (or only up-type) quarks since the corresponding fields are rotated independently. 

After integrating out the heavy diquark, we can write the following effective interactions
\begin{equation}
    \mathcal{L}_{\rm eff}^{d=6} = \frac{\lambda_{\alpha i}\lambda'_{jk}}{M_X^2}\big(\overline{N_\alpha^c} u_R^i \big)\big(\overline{{d_R^c}}^j d_R^k\big) + \frac{\lambda'_{jk}\lambda'_{lm}}{M_X^2}\big(\overline{{d_R^c}}^j d_R^k\big)\big(\overline{{d_R^c}}^l d_R^m\big) + \frac{\lambda_{\alpha i} \lambda_{\beta n}}{M_X^2}\big(\overline{N_\alpha^c} u_R^i \big)\big(\overline{u_R^n} N_\beta \big) +h.c.
    \label{eq:Leff_complete}
\end{equation}

\noindent Apart from generating the baryogenesis mechanism described in the main text, these effective interactions lead to a distinct low-energy phenomenology. As we discussed, provided that $m_N \geq m_p + m_K \approx 1.4 \GeV$ and the operator $\overline{L}HN$ is forbidden in the theory, the proton does not decay. Next, because the diquarks do not directly couple to two quarks of the same generation, tree-level $\Delta F=2$ flavor-changing neutral current (FCNC) processes are avoided. For the same reason, tree-level neutron-antineutron oscillations are also forbidden in the model. At tree level, two main observables put bounds on the couplings of the model. First, tree-level flavor-changing decays can mimic SM decays induced by penguin diagrams. The most constraining ones are $B$ meson decays like $B\rightarrow \phi \pi$ and $B\rightarrow \phi\phi$. Following the calculation done in \cite{Giudice:2011ak}, current limits on the branching ratio of these decays put the following constraints on the diquark couplings
\begin{alignat}{5}
    &\frac{\text{Br}(B^\pm \rightarrow \phi \pi^\pm)}{1.5\times10^{-7}} &&= \left(\frac{|{\lambda'}^*_{ds} \lambda_{sb}'|}{0.035}\right)^2 \left(\frac{2\TeV}{M_X}\right)^4 <1, \text{  at 90\% C.L. \cite{LHCb:2013uvl,ParticleDataGroup:2024cfk}} 
    \\
    &\frac{\text{Br}(B^0 \rightarrow \phi \pi^0)}{1.5\times10^{-7}} &&= \left(\frac{|{\lambda'}^*_{ds} \lambda_{sb}'|}{0.049}\right)^2 \left(\frac{2\TeV}{M_X}\right)^4<1, \text{  at 90\% C.L. \cite{Belle:2012irf,ParticleDataGroup:2024cfk}}  .
\end{alignat}

\noindent Another tree-level process is the $\Delta B=2$ and $\Delta F=2$ dinucleon decay $pp\rightarrow K^+ K^+$ induced by the first operator of \eqref{eq:Leff_complete}. These are constrained by the \textsc{SuperKamiokande} experiment through reactions of the type $O^{16} \rightarrow C^{14} K^+ K^+$ \cite{Litos:2010zra,Takhistov:2016eqm}. Following \cite{Goity:1994dq}, this process put the following limit
\begin{equation}
    |\lambda_{\alpha u}\lambda_{ds}'|\left( \frac{2\TeV}{M_X}\right)^{2} \left( \frac{200\GeV}{m_{N_\alpha}}\right)^{2} < 1.5\times 10^{-6}, \qquad \alpha = 1,2,3.
\end{equation}

\noindent Since we assume the couplings $\lambda'_{ds}$ to be sizable for the production of the new states at the LHC, the dinucleon decay put strong bounds on the up-quark flavor couplings of the diquarks and $N_\alpha$. We therefore assume a hierarchy in which $\lambda_{\alpha u} \ll \lambda_{\alpha c}, \lambda_{\alpha t}$.

At the one-loop level, the most important bounds come from meson oscillations. Refs. \cite{Giudice:2011ak,Han:2010rf,Baldes:2011mh,Pascual-Dias:2020hxo,Han:2023djl} imposed the following constraints from Kaon and B meson oscillations,
\begin{alignat}{5}
    &\label{eq:KKbar}
    \sqrt{|\text{Re}({\lambda'}^*_{db}\lambda'_{sb})^2|} \left(\frac{2\TeV}{M_X}\right) \leq 9.2\times 10^{-2}, \quad &&(K^0 -\overline{K^0})
    \\  
    &\label{eq:BBbar}
    \sqrt{|\text{Re}({\lambda'}^*_{ds}\lambda'_{sb})^2|} \left(\frac{2\TeV}{M_X}\right) \leq 7.2\times 10^{-2}, \quad &&(B_d^0 -\overline{B_d^0}~).
\end{alignat}

\noindent Similarly, due to the $\lambda_{\alpha i}$ couplings, there can also be $D^0$ meson oscillations. The effective Hamiltonian for the transition is given by
\begin{align}
    \mathcal{H}_{\rm eff} = \sum_{\alpha,\beta}  \frac{\lambda_{\alpha u}^* \lambda_{\beta u}\lambda^*_{\alpha c}\lambda_{\beta c} }{16\pi^2 M_X^2} f\left(\frac{m_{N_{\alpha,\beta}}^2}{m_X^2}\right)  (\overline{u} P_R c)(\overline{u} P_R c).
\end{align}

\noindent Since the diquark couplings with $N_{1,2}$ must be small so that they are long-lived, the dominant contribution comes from the $\lambda_{3u}$ and $\lambda_{3c}$ couplings. With only one singlet in the loop the function $f(x)$ is given by,
\begin{equation}
    f(x)=\frac{2(1+x^2 - 2x + x\log x -x^2\log x)}{(1-x)^3} \xrightarrow{x\ll 1} 2 + \mathcal{O}\big(x[1+\ln(x)]\big)
\end{equation}

\noindent Then, following \cite{Golowich:2007ka}, the off-diagonal $D^0$ mass term is
\begin{equation}
    M_{12}=\frac{1}{2 m_D} \langle D^0 | \mathcal{H}_{\rm eff} |\overline{D^0} \rangle = \frac{5}{12} \frac{m_D f_D^2 B_D}{2} \frac{|\text{Re}(\lambda_{3u}^* \lambda_{3c})^2|}{48\pi^2 M_X^2}.
\end{equation}

\noindent where $m_D \simeq 1.9 \GeV$, $f_D\simeq 0.21 \GeV$ are the $D$-meson mass and decay constant and $B_D\approx 0.8$ is the bag factor. Requiring that the mass difference $\Delta m_D = 2M_{12}$ satisfies the experimental bound from \cite{ParticleDataGroup:2024cfk}, $\Delta m_D^{\rm exp} \lesssim 0.94\times 10^{10} \hbar s^{-1}= 6.2 \times 10^{-15} \GeV$, leads to
\begin{equation}
    \sqrt{|\text{Re}({\lambda}^*_{3u}\lambda_{3c})^2|} \left(\frac{2\TeV}{M_X}\right) \leq 2.9\times 10^{-2}, \quad(D^0 -\overline{D^0}).
\end{equation}

\noindent Other one-loop bounds from $\Delta F=1$ processes like $b\rightarrow d \gamma, s \gamma$ and contributions to the chromomagnetic $sd$ operator are weaker than the meson oscillation ones we quote above. For a detailed discussion, we refer to \cite{Giudice:2011ak}.

There are no one-loop contributions for neutron-antineutron oscillations. For $n-\overline{n}$, the absence of one-loop effects can be seen by factorizing the loop diagram into two tree-level ones $(udd \rightarrow \psi_1 \psi_2)\times(\overline{\psi_1 \psi_2} \rightarrow \overline{udd})$ using the optical theorem. The intermediate two-particle state $\psi_1 \psi_2$ must be neutral and carry $\Delta B=2$. This means we need a single Majorana mass insertion, which forces $\psi_1 = N_\alpha$. However, the remaining state $\psi_2$ must also be a fermionic singlet, implying that $\psi_2 = N_\beta$. This leaves no room for a non-vanishing spinor contraction, as one would need two mass insertions to obtain the correct fermionic lines. For the neutron EDM, the one-loop diagrams always involve the modulus-squared of a single coupling, producing no imaginary part. Therefore, there is no one-loop neutron EDM.

At two loops, neutron oscillation is possible but very suppressed. The operator that generates $n-\overline{n}$ is
\begin{equation}
    \mathcal{O}_{n-\bar{n}} = \frac{1}{\Lambda^5_{n\bar{n}}} (u_R d_R d_R)^2.
\end{equation}

\begin{figure}
    \centering
    \includegraphics[width=0.3\linewidth]{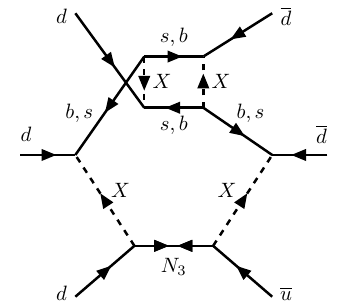}
    \caption{Two-loop contribution to neutron-antineutron oscillations induced by the diquark couplings of the model.}
    \label{fig:nnbar}
\end{figure}

\noindent At two loops, this operator is generated dominantly by the diquark sector as EW flavor-changing processes cannot couple to right-handed currents without light quark mass insertions. One can estimate the effective scale $\Lambda^5_{n\bar{n}}$ by power counting the loop contributions of the type shown in FIG.\ref{fig:nnbar}.
\begin{equation}
    \frac{1}{\Lambda^5_{n\bar{n}}} \sim \sum_\alpha \sum_{i,j,k,l} \frac{|\lambda_{\alpha u}|^2 {\lambda'}^*_{dk}{\lambda}'_{ik} \lambda'_{dl}{\lambda'}^*_{jk}}{256 \pi^4} \frac{m_{N_\alpha}}{M_X^ 6} \sim \frac{|\lambda_{3u}|^2 |\lambda'_{ds}|^4}{256\pi^4}\frac{m_{N_3}}{M_X^ 6}
\end{equation}

\noindent Then the bound for $n-\bar{n}$ oscillations can be obtained as
\begin{equation}
    \tau_{n-\bar{n}} = \frac{\Lambda^5_{n\bar{n}}}{\Lambda_{QCD}^6} \sim 8.6\times 10^{7} s \times\left( \frac{(0.09)^6}{|\lambda_{3u}|^2 |\lambda'_{ds}|^4}\right)\left( \frac{200\GeV}{m_{N_3}}\right)\left( \frac{M_X}{2\TeV}\right)^6
\end{equation}

\noindent Notice that because the coupling $\lambda_{3u}$ is suppressed due to the $pp\rightarrow K^+ K^+$ bound, neutron oscillations are well within the allowed values of $\tau_{n-\bar{n}}^{\rm exp} \approx 8.6\times 10^{7} s$ \cite{ParticleDataGroup:2024cfk}.

Finally, the need for at least two different diquark couplings and a chirality flip suggests that a non-zero neutron EDM contribution only appears at higher loops. In fact, since the only chirality flip in the model occurs through Higgs insertions, there are also no anomalous neutron EDM corrections at two loops. In Ref.\cite{Giudice:2011ak}, it is suggested that the diquark couplings can generate a neutron EDM at three loops, but an estimate of the bounds must be subleading given the other constraints on the couplings and is beyond the scope of this work.

In conclusion, to avoid all low energy bounds it is sufficient to adopt the following parameters,
\begin{alignat}{5}
    &\lambda'_{ds} < 0.3, 
    \qquad &&~
    \lambda'_{db} < 0.3, 
    \qquad &&~ 
    \lambda'_{sb} < 0.11,
    \\
    &\lambda_{\alpha u} < 1.5\times 10^{-6},
    \quad &&~ 
    \lambda_{\alpha c} < 1,
    \quad &&~ 
    \lambda_{\alpha t} < 1, \qquad \alpha = 1,2,3.
\end{alignat}

\end{widetext}

\end{document}